\documentclass{ws-ijmpcs}

\newcommand {\apgt} {\ {\raise-.5ex\hbox{$\buildrel>\over\sim$}}\ }
\newcommand {\aplt} {\ {\raise-.5ex\hbox{$\buildrel<\over\sim$}}\ }

\begin{document}

\markboth{AGUDO ET AL.}
{LOCATION OF THE $\gamma$-RAY EMISSION IN THE JET OF AO~0235+164}

%
\catchline{}{}{}{}{}
%

\title{LOCATION OF THE $\gamma$-RAY FLARING EMISSION IN THE PARSEC-SCALE JET OF THE BL~LAC OBJECT AO~0235+164}

\author{I.~AGUDO$^{1,2}$, A.~P.~MARSCHER$^{2}$, S.~G.~JORSTAD$^{2,3}$,V.~M.~LARIONOV$^{3,4}$, J.~L.~G\'OMEZ$^{1}$, A.~L\"{A}HTEENM\"{A}KI$^{5}$, P.~S.~SMITH$^{6}$, K.~NILSSON$^{7}$, A.~C.~S.~READHEAD$^{8}$, M.~F.~ALLER$^{9}$, J.~HEIDT$^{10}$, M.~GURWELL$^{11}$, C.~THUM$^{12}$, A.~E.~WEHRLE$^{13}$, O.~M.~KURTANIDZE$^{14}$}
\address{$^{1}$Instituto de Astrof\'{\i}sica de Andaluc\'{\i}a (CSIC),  Apartado 3004, E-18080 Granada, Spain\\
$^{2}$IAR, Boston University, 725 Commonwealth Avenue, Boston, MA 02215, USA\\
$^{3}$St. Petersburg University, Universitetskij Pr. 28, Petrodvorets, 198504 St. Petersburg, Russia\\
$^{4}$Isaac Newton Institute of Chile, St. Petersburg Branch, St. Petersburg, Russia\\
$^{5}$Aalto Univ. Mets\"ahovi Radio Observatory, Mets\"ahovintie 114, FIN-02540 Kylm\"al\"a, Finland\\
$^{6}$Steward Observatory, University of Arizona, Tucson, AZ 85721-0065, USA\\
$^{7}$FINCA, University of Turku, V\"ais\"al\"antie 20, FIN-21500 Piikki\"o, Finland\\
$^{8}$CCAA, California Institute of Technology, Mail Code 222, Pasadena, CA 91125, USA\\
$^{9}$DA, University of Michigan, 817 Dennison Building, Ann Arbor, MI 48 109, USA\\
$^{10}$ZAH, Landessternwarte Heidelberg, K\"onigstuhl, 69117 Heidelberg, Germany\\
$^{11}$Harvard-Smithsonian Center for Astrophysics, 60 Garden St., Cambridge, MA 02138, USA\\
$^{12}$IRAM, 300 Rue de la Piscine, 38406 St. Martin dÕH`eres, France\\
$^{13}$Space Science Institute, Boulder, CO 80301, USA\\
$^{14}$Abastumani Observatory, Mt. Kanobili, 0301 Abastumani, Georgia}

\maketitle

\begin{history}
\received{Day Month Year}
\revised{Day Month Year}
\end{history}

\begin{abstract}
We locate the $\gamma$-ray and lower frequency emission in flares of the BL~Lac object AO~0235+164 at $\apgt12$\,pc in the jet of the source from the central engine.
We employ time-dependent multi-spectral-range flux and linear polarization monitoring observations, as well as ultra-high resolution ($\sim0.15$ milliarcsecond) imaging of the jet structure at $\lambda=7$\,mm. 
The time coincidence in the end of 2008 of the propagation of the brightest superluminal feature detected in AO~0235+164 (Qs) with an extreme multi-spectral-range ($\gamma$-ray to radio) outburst, and an extremely high optical and 7\,mm (for Qs) polarization degree provides strong evidence supporting that all these events are related. This is confirmed at high significance by probability arguments and Monte-Carlo simulations. 
These simulations show the unambiguous correlation of the $\gamma$-ray flaring state in the end of 2008 with those in the optical, millimeter, and radio regime, as well as the connection of a prominent X-ray flare in October 2008, and of a series of optical linear polarization peaks, with the set of events in the end of 2008.
The observations are interpreted as the propagation of an extended moving perturbation through a re-collimation structure at the end of the jet's acceleration and collimation zone.

\end{abstract}

\ccode{PACS numbers: 11.25.Hf, 123.1K}

\section{Introduction}
\label{intr}
The physical information about high energy emission mechanisms provided by blazars depends on where those $\gamma$-rays were originated, which is currently subject of debate.\cite{Marscher:2010p14415}
Two main locations of the site of $\gamma$-ray emission in blazars are claimed in the literature.
The first one is a region close to the central black hole (BH), at $\lesssim0.1{\rm{-}}1$\,pc, which is convenient to explain short time-scales of variability of a few hours reported in some $\gamma$-ray observations of blazars.\cite{Ackermann:2010p13506,Tavecchio:2010p14858}
Although short time scales of variability do not necessarily imply short distances to the BH, but only small sizes of the emitting region, this scenario has the advantage that optical-UV photons from the broad line region are available for scattering to $\gamma$-ray energies by relativistic electrons in the jet.
However, locating the $\gamma$-ray emission so close to the central engine implies that the $\gamma$-ray and radio-mm emission sites are far from each other and hence their emission events non-coincident, which seems unlikely for an increasing number of blazars.\cite{Jorstad:2010p11830}\cdash\cite{Agudo:2011p15946}
This problem is overcome if the $\gamma$-rays are emitted from a region much further away from the central engine (at $>>1$\,pc), more concretely where the jet starts to be visible at millimeter wavelengths with VLBI.
Supporting this scenario, Ref.~\refcite{Agudo:2011p14707} unambiguously locate the region of $\gamma$-ray flaring emission at $\apgt14$\,pc from the central engine in the jet of {OJ287} through correlation of the millimeter light curves with the $\gamma$-ray ones and direct ultrahigh-resolution 7\,mm VLBI imaging.
Similar results are obtained by Refs.~\refcite{Marscher:2010p11374} and \refcite{Jorstad:2010p11830} for {PKS 1510$-$089} and {3C~454.3}, respectively.
Ref.~\refcite{Marscher:2010p14415} have recently proposed a model that reconciles both the evidences in favor of the location of $\gamma$-ray flare emitting regions $>>1$\,pc from the BH with the ``few-hour'' time scales of $\gamma$-ray variability through a turbulence model with frequency-dependent filling factor.
Following Ref.~\refcite{Agudo:2011p14707} for OJ287, we use here time series of total flux and polarization multiwavelength observations (including monthly VLBI images) to investigate the location of the $\gamma$-ray emitting region in the BL~Lac object AO~0235+164  ($z=0.94$).

\section{Observations}
\label{obs}
\begin{figure}[pb]
\centerline{\psfig{file=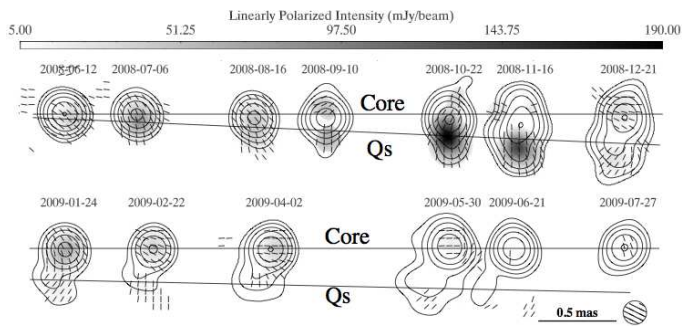,width=9.cm}}
\vspace*{8pt}
\caption{7\,mm VLBA image-sequence of AO~0235+164 convolved with a $\rm{FWHM}=0.15$\,mas circular-Gaussian beam. Contour levels represent the observed total intensity (9 levels from $0.4$ to 90.0\,\% of total intensity peak$=4.93$\,mJy/beam), the grey scale symbolizes polarized intensity, and superimposed sticks show the orientation of $\chi$.}
   \label{maps}
\end{figure}

Our polarimetric monitoring observations of AO~0235+164 (Figs.~\ref{maps}-\ref{pol}) include (1) 7\,mm images with the Very Long Baseline Array (VLBA) from the Boston University monthly blazar-monitoring program, (2) 3.5\,mm observations with the IRAM 30\,m Telescope, and (3) optical measurements from several observatories.  
These include Calar Alto (2.2\,m Telescope), Steward (2.3 and 1.54\,m Telescopes), Lowell (1.83\,m Perkins Telescope), San Pedro M\'{a}rtir (0.84\,m Telescope), Crimean Astrophysical (0.7\,m Telescope), and the St. Petersburg State University (0.4\,m Telescope) observatories. 
Our total flux light curves (Fig.~\ref{tflux}) include data from the \emph{Fermi} $\gamma$-ray (0.1-200\,GeV) and \emph{Swift} X-ray (2.4-10\,keV) observatories available from the archives of these missions, from \emph{RXTE} (2.4-10\,keV), from the Tuorla Blazar Monitoring Program, the Yale University SMARTS program and Maria Mitchell Observatory in the optical $R$-band, by the Submillimeter Array (SMA) at 1.3\,mm and 850\,$\mu$m, by the IRAM 30\,m Telescope at 1.3\,mm, by the Mets\"{a}hovi 14\,m Telescope at 8\,mm, and by the Owens Valley Radio Observatory and the University of Michigan Radio Astronomy Observatory 26\,m Telescope at 2\,cm.
For the data reduction we followed the procedures used in Ref.~\refcite{Agudo:2011p15946} (see also \refcite{Jorstad:2010p11830,Marscher:2010p11374,Agudo:2006p203,Agudo:2010p12104,Larionov:2008p338}).

\section{Superluminal Jet Ejection from the 2008 Millimeter Flare}
\label{ejection}

Our 7\,mm VLBA maps of AO~0235+164 (Fig.~\ref{maps}) show a compact total intensity distribution that is fitted by one or two circular Gaussian components in most of the observing epochs, when preserving consistency from epoch to epoch and flat residual maps. 
We identify the core of emission in our images with the innermost visible mm emitting region in the jet.
Between mid-2008 and mid-2009, the jet structure also shows a second emission region (Qs, the brightest 7\,mm jet feature detected so far) that propagates at a superluminal apparent speed $<{\beta_{\rm{app}}}>=(12.6\pm1.2)\,c$.

\begin{figure}[pb]
\centerline{\psfig{file=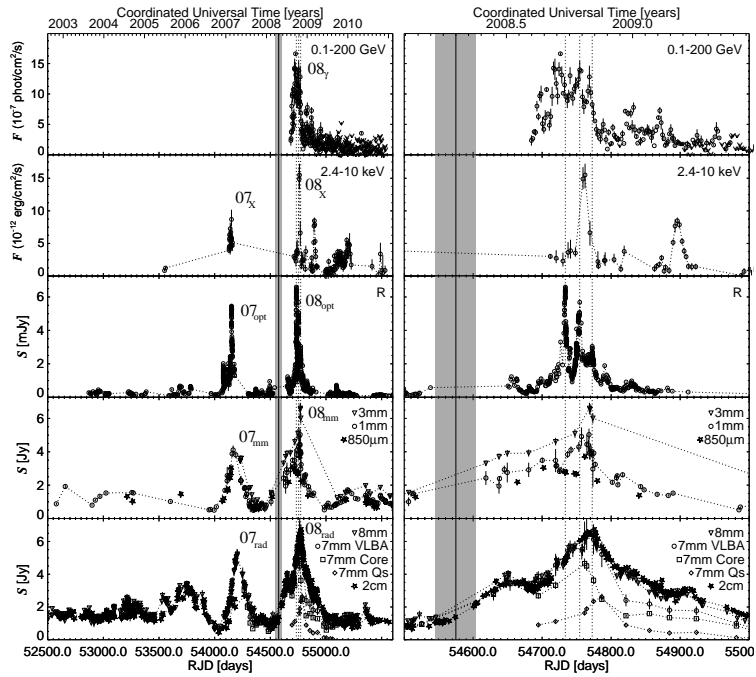,width=10.cm}}
\vspace*{8pt}
\caption{\emph{Left:} Long term light curves of AO~0235+164 from $\gamma$-ray to millimeter frequencies. Vertical dotted lines symbolize the time of the three optical peaks discussed in the text, whereas the grey area represents the ejection time range of feature Qs. RJD = Julian Date $- 2400000.0$. \emph{Right:} Same as Fig~\ref{tflux} but zoomed into RJD=[54500,55000].}
  \label{tflux}
\end{figure}

The ejection of Qs, that we estimate in $2008.30\pm0.08$, coincides with the start of an extreme millimeter outburst (${08}_{\rm{mm}}$) started in early 2008, and peaking (at $\sim6.5$\,Jy at 3\,mm) on October 10$^{\rm{th}}$, 2008 (see Fig.~\ref{tflux}).
Figure~\ref{tflux} shows that ${08}_{\rm{rad}}$ and ${08}_{\rm{mm}}$ radio and millimeter outbursts are produced by the contribution of both the VLBI core and Qs, that peak around October 20$^{\rm{th}}$ and November 16$^{\rm{th}}$ 2008, respectively.
Their coherent -although delayed- co-evolution suggests that the disturbance responsible for the ejection of Qs is wide and extends from the location of the core to Qs in the frame of the observer via light travel time delay effects.\cite{1997ApJ...482L..33G,Agudo:2001p460}
Qs is the only strong superluminal ejection during our observations and the ${08}_{\rm{rad}}$ and ${08}_{\rm{mm}}$ flares are the only ones starting essentially after such ejection, which allow us to unambiguously relate both kinds of events.

\section{Flaring Correlation from $\gamma$-rays to Radio Wavelengths}
\label{corr}
Figure~\ref{tflux} reveals that the ${08}_{\rm{rad}}$ and ${08}_{\rm{mm}}$ flares were accompanied by sharp optical, X-ray, and $\gamma$-ray counterparts (${08}_{\rm{opt}}$, ${08}_{\rm{X}}$, and  ${08}_{\gamma}$ flares, respectively). 
Our formal correlation study of the light curves in Fig.~\ref{tflux}\cite{Agudo:2011p15946} confirms the correlation of the $\gamma$-ray variability with the one at 2\,cm, 8\,mm, 1\,mm, and optical with confidence $>99.7$\,\%.
Even the integrated flux evolution in the 7\,mm VLBI-core region is also correlated with the $\gamma$-ray light curve at $>$99.7\,\% confidence.
The optical linear polarization degree ($p_{\rm{opt}}$) evolution and X-ray light curve are also correlated with the optical $R$-band, 1\,mm, and 2\,cm light curves at $>99.7$\,\% confidence, hence pointing out that the extreme flaring activity shown in our light curves is related at all wavebands from radio to $\gamma$-rays.

Note however that there is not a common variability pattern at all spectral ranges on short time scales ($\lesssim2$~months), although there is clear correlation on the long time scales (of $\sim$years), hence reflecting different variability properties on short and long time scales.

\begin{figure}[pb]
\centerline{\psfig{file=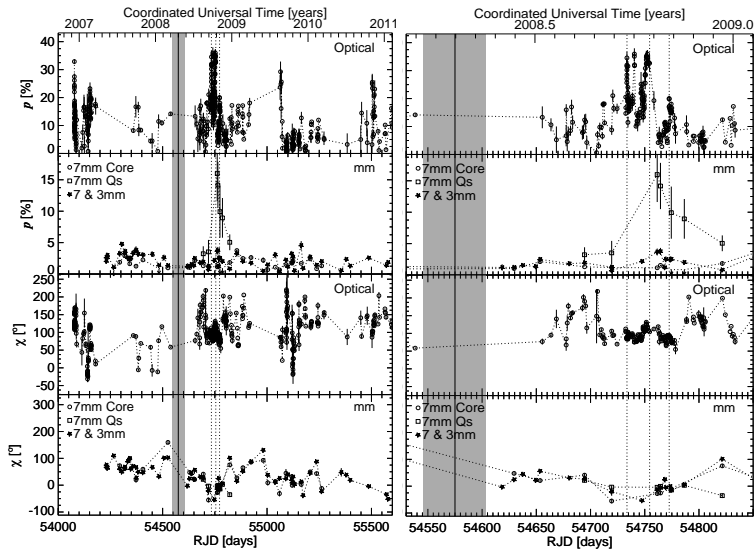,width=10.cm}}
\vspace*{8pt}
\caption{\emph{Left:} Long term optical and millimeter linear polarization evolution of AO~0235+164 in the RJD=[54000,55600] range. \emph{Right:} Same as Fig.~\ref{tflux} but zoomed into RJD=[54530,54850].}
   \label{pol}
\end{figure}

\section{Relation with the Linear Polarization Variability}
\label{polvar}

Figure \ref{pol} show extremely high optical polarization fraction ($p_{\rm{opt}}\gtrsim30$\,\%) and $p_{\rm{opt}}$ variability during the sharp ${08}_{\rm{opt}}$ optical peaks.
Whereas the integrated millimeter linear-polarization-degree ($p_{\rm{mm}}$), and the one of the core are observed to lie at moderate values $\lesssim5$\,\%, for Qs, $p_{\rm{mm,Qs}}$ shows an extremely high peak at $\sim16$\,\% close to the time of the second sharp optical sub-flare.
Such an extraordinary $p_{\rm{mm,Qs}}$ peak -by far the brightest superluminal feature ever detected in AO~0235+164- has never been observed in other VLBI jet feature in the source.
Its coincidence with the high optical flux and $p_{\rm{opt}}$ state, and with the flaring states at the remaining spectral ranges (including the 7\,mm-VLBI-core one), evidences that the ejection and propagation of Qs in AO~0235+164's millimeter jet is physically tied to the contemporaneous multi-spectral-range flaring state and the extreme $p_{\rm{opt}}$ activity reported here.

\section{Discussion and Conclusions}
The time coincidence of the ejection and propagation of Qs -the brightest superluminal feature reported in AO~0235+164 so far- with the extreme $\gamma$-ray to radio outbursts in 2008 reported here and the extremely high $p_{\rm{opt}}$ and $p_{\rm{mm,Qs}}$, provides extremely strong evidence in favor of the physical relation of all these events. 
This is confirmed by simple probability arguments\cite{Agudo:2011p15946} -that give $p_{\gamma,{\rm{opt}},{\rm{mm}}}=5\times10^{-4}$ for observing, by chance, a contemporaneous $\gamma$-ray, an optical, and a radio-millimeter flare- and by our formal correlation study.\cite{Agudo:2011p15946}
The latter shows the unambiguous correlation of the $\gamma$-ray outburst in the end of 2008 with those in the optical, millimeter (including the 7\,mm VLBI core), and radio spectral ranges.
The location of the 7\,mm VLBI-core $\gtrsim12$\,pc from the jet vertex in AO~0235+164\cite{Agudo:2011p15946}, together with the high-confidence correlation of the 7\,mm core flare in 2008 with the multi-wavelength flare at all other spectral ranges\cite{Agudo:2011p15946}, evidences that the emission from the flares at these different spectral ranges was also located $\gtrsim12$\,pc from the central engine.

We identify the 7\,mm VLBI-core as a stationary re-collimation structure at the end of the acceleration and collimation zone of the jet's acceleration and collimation zone.\cite{Marscher:2010p11374,Marscher:2008p15675}
Qs superluminal feature is rather related to a moving plane-perpendicular shock, given the extremely high $p_{\rm{mm,Qs}}$, and $\chi_{\rm{mm}}^{\rm{Qs}}$ parallel to Qs's direction of propagation.
The core shows a flux evolution closely tied to the Qs one, and its light curve is correlated at high confidence with those at $\gamma$-rays, optical, and millimeter frequencies.
This suggests that Qs is the head of an extended structure stretched by light travel time-delays in the observer's frame, as e.g. the front-back structure reported by Ref.~\refcite{Aloy:2003p350}.

Multi-zone SSC scenarios, e.g. as those involving emission from different turbulence cells\cite{Marscher:2010p14415}, are likely for the case of AO~0235+164 given its good multi-spectral-range correlation on the long time scales, but poorer short time-scale correlation (see also Ref.~\refcite{Agudo:2011p15946}). 

\section{Acknowledgements}
This research was partly funded by NASA grants NNX08AJ64G, NNX08AU02G, NNX08AV61G, and NNX08AV65G, NSF grant AST-0907893, NRAO award GSSP07-0009; RFBR grant 09-02-00092; MICIIN grant AYA2010-14844; CEIC grant P09-FQM-4784; and GNSF grant ST08/4-404.
The VLBA is an instrument of the NRAO, a facility of the NSF operated under cooperative agreement by AUI. 
The PRISM camera at Lowell Observatory was developed by Janes et al. 
The Calar Alto Observatory is jointly operated by MPIA and IAA-CSIC. 
The IRAM 30\,m Telescope is supported by INSU/CNRS, MPG, and IGN.
The SMA is a joint project between the SAO and the Academia Sinica. 


\end{document}